\documentclass[iop]{emulateapj}
\shorttitle{A massive neutron star with a giant secondary}
\shortauthors{Strader \etal~}
\def\etal{{et al.}}
\def\kms{\,km~s$^{-1}$}

\def\arcsec{\char'175 }

\def\hub{\ifmmode H_\circ\else H$_\circ$\fi}
\usepackage{subfigure}
\usepackage{amsmath}
\usepackage{graphics}
\usepackage{hyperref}
\usepackage{breakurl} 

\def\ltsima{$\; \buildrel < \over \sim \;$}
\def\simlt{\lower.5ex\hbox{\ltsima}} 
\def\gtsima{$\; \buildrel > \over \sim \;$}
\def\simgt{\lower.5ex\hbox{\gtsima}} 
\def\arcsec{\hbox{$^{\prime\prime}$}}
\def\deg{\hbox{$^\circ$}}

\def\gray{$\gamma$-ray}

\def\Fermi{\textit{Fermi}}

\def\Chandra{\textit{Chandra}}

\def\zJ0523{1FGL~J0523.5$-$2529}
\def\aJ1417{1FGL~J1417.7$-$4407}

\begin{document}

\title{

1FGL~J1417.7$-$4407: A likely $\gamma$-ray bright binary with a massive neutron star and a giant secondary}
\author{
Jay Strader\altaffilmark{1},
Laura Chomiuk\altaffilmark{1},
C.~C.~Cheung\altaffilmark{2},
David J.~Sand\altaffilmark{3},
Davide Donato\altaffilmark{4,5},
Robin H.~D.~Corbet\altaffilmark{4},
Dana Koeppe\altaffilmark{1},
Philip~G.~Edwards\altaffilmark{6},
Jamie~Stevens\altaffilmark{6}, 
Leonid Petrov\altaffilmark{7},
Ricardo Salinas\altaffilmark{1},
Mark Peacock\altaffilmark{1},
Thomas Finzell\altaffilmark{1},
Daniel E.~Reichart\altaffilmark{8},
Joshua B.~Haislip\altaffilmark{8}}

\altaffiltext{1}{Department of Physics and Astronomy, Michigan State University, East Lansing, MI 48824, USA}
\altaffiltext{2}{Space Science Division, Naval Research Laboratory, Washington, DC 20375, USA}
\altaffiltext{3}{Department of Physics, Texas Tech University, Box 41051, Lubbock, TX 79409, USA}
\altaffiltext{4}{Center for Research and Exploration in Space Science and Technology (CRESST) and NASA Goddard Space Flight Center, Greenbelt, MD 20771, USA}
\altaffiltext{5}{Department of Physics and Department of Astronomy, University of Maryland, College Park, MD 20742, USA}
\altaffiltext{6}{CSIRO Astronomy and Space Science, PO Box 76, Epping, NSW 1710, Australia}
\altaffiltext{7}{Astrogeo Center, Falls Church, VA 22043, USA}
\altaffiltext{8}{Department of Physics and Astronomy, University of North Carolina at Chapel Hill, Chapel Hill, NC 27599, USA}

\begin{abstract}

We present multiwavelength observations of the persistent \Fermi-LAT unidentified $\gamma$-ray source \aJ1417, showing it is likely to be associated with a newly discovered X-ray binary containing a massive neutron star (nearly $2 M_{\odot}$) and a $\sim 0.35 M_{\odot}$ giant secondary with a 5.4 day period. SOAR optical spectroscopy at a range of orbital phases reveals variable double-peaked H$\alpha$ emission, consistent with the presence of an accretion disk. The lack of radio emission and evidence for a disk suggests the $\gamma$-ray emission is unlikely to originate in a pulsar magnetosphere, but could instead be associated with a pulsar wind, relativistic jet, or could be due to synchrotron self-Compton at the disk--magnetosphere boundary. Assuming a wind or jet, the high ratio of $\gamma$-ray to X-ray luminosity ($\sim 20$) suggests efficient production of $\gamma$-rays, perhaps due to the giant companion. The system appears to be a low-mass X-ray binary that has not yet completed the pulsar recycling process. This system is a good candidate to monitor for a future transition between accretion-powered and rotational-powered states, but in the context of a giant secondary.

\end{abstract}
 

\section{Introduction\label{section-intro}}

Millisecond pulsars are thought to have been recycled to short periods through accretion from either a main sequence or giant secondary. Despite the expectation that this mass transfer process should be lengthy, most known field millisecond pulsars have degenerate white dwarf companions that represent the end stage of the recycling process (see the review of Tauris \& van den Heuvel 2006). This suggests the radio emission is quenched during even low-level accretion onto the neutron star, making the systems undetectable as pulsars.

The discovery of PSR J1023+0038 represented the first compelling evidence that some millisecond pulsars actively switch between a low state with radio pulsations and a high state with an accretion disk (Archibald et al.~2009), which in this system is due to Roche lobe overflow of its $\sim 0.2 M_{\odot}$ bloated main sequence companion (Archibald et al.~2013; Bogdanov et al.~2015). Two additional similar systems have been discovered, both with main sequence companions (Papitto et al.~2013; Bassa et al.~2014; Roy et al.~2015).

These transitional systems belong to a larger class of objects known as ``redbacks", binary systems in which the wind of the pulsar primary ablates a main sequence companion of $\gtrsim 0.1 M_{\odot}$, causing radio eclipses for a substantial fraction of the orbit. The \emph{Fermi} Large Area Telescope (LAT) has been a boon for the discovery of redbacks: many of these are $\gamma$-ray sources that had not previously been detected as pulsars, possibly because of the presence of ionized material in the system (Roberts 2013). 

As part of an ongoing survey of unidentified \emph{Fermi}-LAT sources (Strader et al.~2014), this paper reports observations of a faint $\gamma$-ray source from the 1FGL catalog: \aJ1417 (Abdo \etal~2010). Using X-ray, radio, and optical observations including photometry and spectroscopy, we find that \aJ1417 is likely a Galactic compact binary with a massive neutron star primary and a giant secondary in a relatively wide orbit.

\section{Observations\label{sec-observations}}

\subsection{Fermi-LAT Source}

The $\gamma$-ray source was first catalogued as 1FGL J1417.7$-$4407 \citep{1fgl}, detected in 11 months of LAT all-sky survey data. In the recently released 3FGL catalog \citep{3fgl}, derived from 4 years of LAT observations, it is listed as 3FGL J1417.5$-$4402. The 3FGL 95\% error ellipse ($3.7' \times 3.5'$ at position angle 17\deg), centered at epoch J2000 (R.A., Dec.) = (214.377\deg, $-$44.043\deg), is $\sim9\times$ smaller in area than the 1FGL error ellipse. The 3FGL source is characterized by a single power-law spectrum with an index of $2.37\pm0.08$, with no significant evidence for a curved spectrum. The 0.1--100 GeV flux corresponds to a luminosity of $(2.8\pm0.3) \times 10^{34} \, (d/4.4 \, {\rm kpc})^2$ erg s$^{-1}$. The 3FGL catalog light curve appears superficially variable, but the variability index is 55 (see Nolan et al.~2012 for a definition), below the 99\% threshold of $\sim 72$ necessary to classify the variability as ``probable" in the 3FGL catalog. Therefore we assess the $\gamma$-ray variability as uncertain with present data (though there is evidence for long-term X-ray variability; see below).

\subsection{X-ray Observations}

\aJ1417\ was observed with \Chandra\ as part of a Cycle 12 program (PI: C.~Ricci) targeting six unidentified 1FGL \emph{Fermi}-LAT sources. The $\sim 2$ ksec exposure was obtained on 2011 Feb 10, and had placed the target on the ACIS-I detector ($\sim 17' \times 17'$ field of view) to ensure the entire \aJ1417\ error ellipse was covered. The XASSIST processed catalog source, X141730.55$-$440257.6 \citep[J2000 position based name; see][]{pta03}, was the only X-ray source detected in the \Chandra\ exposure within the 3FGL \emph{Fermi}-LAT \gray\ error ellipse (\S~2.1), being located essentially at its centroid ($\sim 0.4$\arcmin\ offset).

We downloaded the \Chandra\ data from the archive and analyzed the data using the CIAO software v.4.7 \citep{fru06} and the most recent CALDB v4.6.5. The level 1 (evt1) files were reprocessed using standard procedures in CIAO to generate new evt2 files. We found no background flares during the observation, thus the net exposure was 2051.2 s from 09:43:10.032 to 10:17:21.232 UTC ($\phi=0.94$ using the ephemeris from \S 3). For spectral analysis, we extracted photons with energies $0.5-7$ keV within a circle with radius 5\arcsec\ centered on the X-ray position, with the background determined from an annulus from 15\arcsec\ to 160\arcsec, giving a net count rate of $26.1 \pm 3.6$ cts ksec$^{-1}$ (53 photons).

We found that an absorbed single power-law fit to the data could not be distinguished from an absorbed blackbody spectrum (cstat/dof = 192.7/444 and 192.2/444, respectively). The best-fit power-law had a photon index $\Gamma = 1.32 \pm 0.40$ and an observed  $0.5-7$ keV flux of $(5.6 \pm 1.4) \times 10^{-13}$ erg cm$^{-2}$ s$^{-1}$ ($90\%$ confidence), while the blackbody had $kT = 0.78^{+0.16}_{-0.12}$ keV with flux $(4.8^{+1.4}_{-1.0})\times 10^{-13}$ erg cm$^{-2}$ s$^{-1}$. These values were derived assuming the Galactic absorption, $N_H = 6.54 \times 10^{20}$ cm$^{-2}$ \citep{kal05}, and gave unabsorbed fluxes $5.9\times10^{-13}$ and $5.0\times10^{-13}$ erg cm$^{-2}$ s$^{-1}$ for the respective models. There were hints of additional absorption but when left to vary, the errors are compatible with zero, thus only the Galactic absorption fixed values are presented. The power-law flux is equivalent to a luminosity of $(1.4\pm0.4) \times 10^{33} (d/4.4 \, \textrm{kpc})^2$ erg s$^{-1}$, where the uncertainty is at 90\% confidence as above.

To gauge possible variability, we divided the counts into two bins of 1025.6 s each and found net count rates of $36.5 \pm 6.0$ (37 photons) and $15.7 \pm 4.0$ (16 photons) counts ksec$^{-1}$ in the respective bins. Assuming the power-law model, the unabsorbed $0.5-7$ keV fluxes are  $(6.9 \pm 1.9) \times 10^{-13}$ and $4.0^{+1.6}_{-1.7} \times 10^{-13}$ erg cm$^{-2}$ s$^{-1}$, respectively, and are consistent within the joint uncertainties. Due to the low count rate and unfortunate location of the source over a node in an ACIS-I chip, these data are not well-suited to provide constraints on the shorter-term
X-ray variability observed for known transitional millisecond pulsars (e.g., Bogdanov et al.~2015).

We note that ROSAT catalogued an X-ray source, 2RXP J141731.0-440253 \citep[observation on 1997 Feb 09;][]{ros00} that is only 6.5\arcsec\ offset from the \emph{Chandra} source position, thus well within the typical positional uncertainty for ROSAT sources \citep{vog99}. The ROSAT PSPC count rate of this source is $42.4\pm8.0$ counts ksec$^{-1}$ over 0.1--2.4 keV, which is equivalent to observed 0.5--7 keV fluxes of $1.7 \times 10^{-12}$ erg cm$^{-2}$ s$^{-1}$ and $2.0 \times 10^{-12}$ erg cm$^{-2}$ s$^{-1}$ assuming the best power-law and blackbody models derived from the \Chandra\ data. These are a factor of 3 or more brighter than the corresponding \emph{Chandra} fluxes, thus there is evidence for long-term X-ray variability of the source. The source has not been detected by all-sky X-ray monitors such
as \emph{Swift}/BAT (Baumgartner et al.~2013) or \emph{MAXI} (Hiroi et al.~2013), suggesting it has not undergone an outburst in the recent era of these missions.

\subsection{Radio Observations}


Petrov \etal~(2013) presented radio continuum observations at 5.5 and 9.0 GHz with the Australian Telescope Compact Array of a number of unidentified \emph{Fermi}-LAT sources. During these observations, obtained on 2012 Sep 20, no radio emission was detected at the position of the X-ray and optical counterpart discussed in this paper. The only radio source found within the 3FGL ellipse is a faint flat-spectrum radio source ($2.6 \pm 0.3$ mJy at 5.5 GHz; $\alpha = +0.2; S_{\nu} \propto \nu^{\alpha}$) at R.A.~= 14:17:16.39, Dec.~= $-$44:04:39.4 offset by $\sim 3.1$\arcmin\ from the X-ray/optical position. 

This field was re-observed using the 6 km configuration on 2013 Feb 4, with two 15.5 min pointings simultaneously observed again at 5.5 and 9 GHz, each with a bandwidth of 2 GHz. Again, no radio emission associated with the X-ray/optical source was detected, with a flux density upper limit of 0.15 mJy at both frequencies. The field radio source was found to be significantly variable with a 5.5 GHz flux density of $1.43 \pm 0.07$ mJy ($\alpha$ = --0.2) in the 2013 observation; it is not detected in the Chandra data with a 0.5--7 keV X-ray upper limit of $<1.5 \times 10^{-14}$ erg cm$^{-2}$ s$^{-1}$. The results of Schinzel et al.~(2015) imply a probability of 15\% to find an unassociated 2 mJy compact radio source within $3.1'$ of a \emph{Fermi}-LAT source. In addition, extragalactic high-latitude \emph{Fermi}-LAT sources are mostly blazars with radio flux densities 10--100 times higher than the 2 mJy measured for this source (Ackermann et al.~2015). Thus it is reasonable to conclude this radio source is unrelated to the $\gamma$-ray source.



\subsection{The Optical and Near-IR Counterpart}

Using the USNO B1.0 catalog \citep{mon03}, we found an optical counterpart within 0.63\arcsec\ of the X-ray source, located at a J2000 sexigesimal position of (R.A., Dec.) = 
(14:17:30.604, --44:02:57.37). This source has optical magnitudes $B2=16.90$, $R2=15.79$, and $I=14.98$ mag. It is coincident with a 2MASS point source with $J=14.17\pm0.03$, $H = 13.50\pm0.03$, and $K = 13.41\pm0.04$ (Cutri \etal~2003). It is also present in the UCAC4 astrometric catalog (Zacharias et al.~2013), with a ($\mu_{\alpha} \textrm{cos} \, \delta$, $\mu_{\delta}$) proper motion of (--$8.8\pm5.0$, --$3.1\pm4.7$) mas yr$^{-1}$.
 
\subsubsection{Catalina Sky Survey}

We obtained archived photometric observations of the USNO B1.0 source from the Catalina Sky Survey \citep[CSS;][]{dra09} and its associated southern Siding Spring Survey (SSS). These surveys take unfiltered images in sequences of four 30-sec exposures typically reaching $V$-equivalent magnitudes of  $\sim 19$--20 mag. We found the source catalogued in SSS as J141730.6-440257, with 195 photometric measurements from 2005 Aug 15 to 2013 May 7. Four of these had large uncertainties ($> 0.15$ mag) and were removed, leaving a total of 191 measurements.  The median magnitude is $V_{\rm equiv} = 15.85$ and median uncertainty 0.09 mag. There is no evidence for a change in the mean brightness of the source over the range covered by these data.

Using the Lomb-Scargle periodogram as implemented in {\tt R} (Ruf 1999), we searched for the most likely photometric period in the SSS photometry. This period is 2.69 d, but it turns out that this is an alias of the real period determined via spectroscopy (\S 3.1). This period is twice as long: $5.3737\pm0.0003$ d, with the uncertainty estimated via bootstrap. 

\subsubsection{PROMPT}

We obtained time series photometry of the optical source in $BVR$ with the PROMPT-5 telescope (Reichart \etal~2005) at Cerro Tololo International Observatory between 2014 June 15 and 2014 August 29. The data were reduced in the standard manner, and aperture photometry performed to obtain instrumental magnitudes for the target source differentially with respect to five comparison stars in the field. These magnitudes were calibrated using observations of Landolt (1992) standard star fields observed over a range of airmasses on most nights, with zeropoints and extinction coefficients determined on photometric nights. 

\subsection{Optical Spectroscopy}

\subsubsection{SOAR Spectroscopic Monitoring}

We obtained two low-resolution spectra of the candidate optical counterpart to \aJ1417 on 2013 Jan 15 using the Goodman High-Throughput Spectrograph (Clemens \etal~2004) on the SOAR 4.1-m telescope using a 1.03\arcsec\ slit and a 600 l mm$^{-1}$ grating (resolution 3 \AA). The spectrum appears to be that of a late G or early K star (see Figure 1), with the exception of a bright H$\alpha$ emission line with an equivalent width of $\sim$ 8 \AA. The H$\alpha$ emission line has a double-peaked morphology, with the centroids of the blue and red peaks each offset by $\sim 120$\kms\ with respect to the photospheric velocity of the star. This emission is consistent with the presence of an accretion disk, and considering its X-ray detection, we concluded this source was likely to be the true optical counterpart of the \emph{Fermi} source.

\begin{figure}[ht]
\includegraphics[width=3.3in]{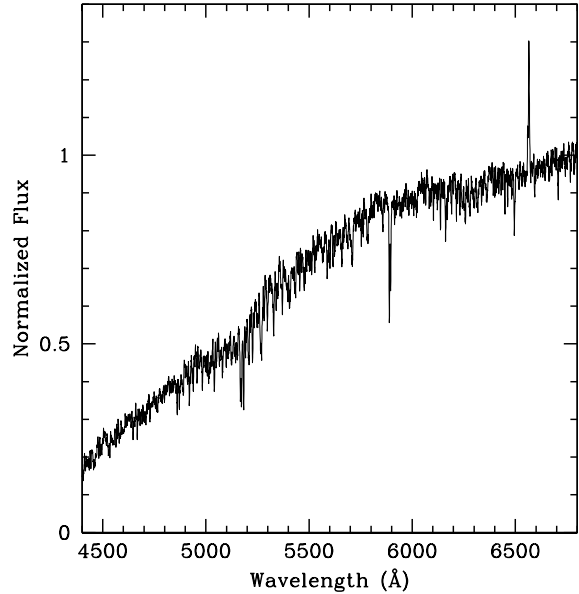}
\caption{Low-resolution SOAR spectrum from 2013 Jan 15 ($\phi = 0.12$) showing the late-G/early-K spectral type of the secondary and the clear H$\alpha$ emission.}
\end{figure}

The following season we began spectroscopic monitoring of this optical counterpart, taking spectra in 14 epochs from 2014 Jan 11 to 2015 Feb 18. Most observations were taken with a 1.03\arcsec\ slit and a 1200 l mm$^{-1}$ grating centered at 6100 \AA, yielding a resolution of 1.7 \AA\ and wavelength coverage of $\sim 5375$--6640 \AA. This enabled measurements of the photospheric radial velocity as well as monitoring of the H$\alpha$ emission. At a few epochs, due to the unavailability of the 1200 l mm$^{-1}$ grating, we instead used a 2100 l mm$^{-1}$ grating, which had a resolution of 0.9 \AA\ and wavelength coverage $\sim 4960$--5600 \AA. The exposure time for each of the 1200 and 2100 l mm$^{-1}$ spectra was 10 minutes. Wavelength calibration was performed using FeAr arcs taken after each set of three exposures.

All spectra were reduced in the usual manner, with optimal extraction and wavelength calibration using the FeAr arcs. To correct for flexure, we cross-correlated the sky spectrum for each exposure with a master sky spectrum over the wavelength range 6275--6375 \AA, which is rich in sky lines (for the 2100 l mm$^{-1}$ data, we used the 5577.34 \AA\ line).

We derived a barycentric radial velocity of each spectrum through cross-correlation with a set of bright stars taken with the same setup, excluding the region around H$\alpha$ in all cases. The median formal uncertainty on the radial velocities is 5.2 \kms.

\subsubsection{High-Resolution VLT Spectroscopy}

In cross-correlation of the SOAR spectra with bright stars of similar spectra type, we found marginal evidence for broadening of the lines, as would be expected if the star was rapidly rotating due to tidal locking with the primary. However, the resolution of the SOAR spectra was too low to precisely measure the line-broadening.

We obtained a high-resolution spectrum of the optical counterpart to \aJ1417 on 2014 Aug 20 using UVES on UT2 of the Very Large Telescope (VLT). Three exposures were taken, each 708 sec in length. The wavelength coverage was $\sim 4800$--5750 \AA\ and 5850--6800 \AA\ with a resolution of $R\sim43000$. Using the same instrumental setup, we also obtained a spectrum of the bright K2III star HD 132096. All spectra were reduced using the standard UVES pipeline. To estimate the projected rotational velocity ($V_{rot}$ sin $i$) of the optical counterpart, we divided the spectrum of HD 132096 into 100 \AA\ chunks, convolved these with a set of kernels reflecting a range of $V_{rot}$ sin $i$ values (including limb darkening), and cross-correlated these with the unbroadened spectra (Strader \etal~2014). We then fit the relation between the FWHM of the cross-correlation peak and the input value of $V_{rot}$ sin $i$. Finally we cross-correlated the spectrum of the optical counterpart to \aJ1417 with that of the standard star in each 100 \AA\ region, using the dispersion among the measurements as an estimate of the random uncertainty in the calculation.

The value of $V_{rot}$ sin $i$ we obtain is $33.6\pm0.7$ \kms. We note that this uncertainty only reflects the random uncertainty in the calculation, and that the systematic uncertainties are likely to be larger.

\section{Results}

\subsection{Keplerian Orbit Fitting and Mass Ratio}

As in our previous study of the compact binary 1FGL~J0523.5$-$2529 (Strader \etal~2014), we performed a Keplerian fit to the 47 radial velocities using the IDL package {\tt BOOTTRAN} (Wang et al.~2012) after correcting the observation epochs to Barycentric Julian Date (BJD) on the Barycentric Dynamical Time (TDB) system (Eastman \etal~2010). Initially we fixed the eccentricity $e=0$ and fit for the period $P$, semi-amplitude $K_2$, systemic velocity $v_{sys}$ and BJD at superior conjunction ($T_{0.5}$; this is when the secondary is behind the primary). We found an excellent fit ($\chi^2$ = 42/47 d.o.f.; rms 5.1 \kms), with orbital elements $P = 5.37385\pm0.00035$ d; $K_2 = 115.7\pm1.1$ \kms; $v_{sys} = -15.3\pm0.9$ \kms; $T_{0.5} = 2457067.605\pm0.024$ d (given as the BJD preceding the final spectroscopic dataset). A fit with the eccentricity free yields $e = 0.01\pm0.01$, so we find no significant evidence for a non-zero eccentricity. The spectroscopic and photometric periods are consistent to within about 15 s; for the remainder of the paper, we adopt the spectroscopic period. The phased radial velocity curve is shown in Figure 2.

Using the standard formula for close binaries (Casares 2001), the mass ratio $q = M_2/M_1$ is directly determined by the rotational velocity and semi-amplitude of the secondary:
$V_{rot} \, \textrm{sin} \, i = 0.462\, K_2\, q^{1/3}\, (1+q)^{2/3}$. Using the values above, this gives $q=0.179\pm0.010$.

\begin{figure}[ht]
\includegraphics[width=3.3in]{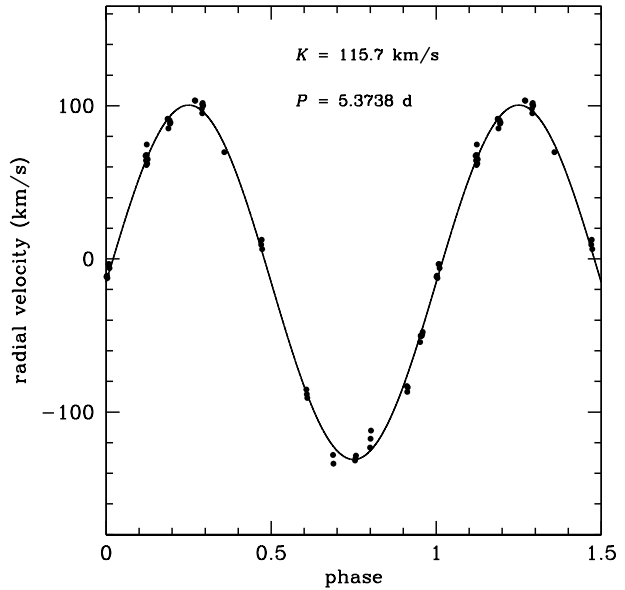}
\caption{Phased radial velocities for the optical counterpart to \aJ1417, with the listed Keplerian fit overplotted.}
\end{figure}

\subsection{Properties of the Secondary}

Given the evidence for an accretion disk and ellipsoidal variations, if we assume the secondary is just filling its Roche lobe, then the standard formula (Eggleton 1983) immediately yields the mean density given the known mass ratio and period. This density is $\bar{\rho} = 0.0068$ g cm$^{-3}$. Thus the secondary is clearly a giant.

The visual impression of the spectrum is that of a late-G or early-K spectral type. To quantify this, we used the package {\tt MKCLASS} (Gray \& Corbally 2014), which can perform MK classification on flux-calibrated spectra. We used two low-resolution spectra taken in 2014 August at phases of $\phi \sim 0.19$ and 0.61. We obtained identical best-fit spectral types of G9 for both spectra, and a metallicity of  [Fe/H] = --0.65. Both of these estimates should be taken as uncertain given the low resolution of the spectra. Unfortunately, the high-resolution UVES spectrum is not suitable for detailed analysis given its modest S/N and the rotational broadening.

We can also study the ``night side" properties of the secondary with photometry. At this $\phi = 0$, we estimate $B\sim17.18$ and $V\sim16.22$ from the PROMPT photometry (see Figure 3; these values are approximate owing to the incomplete phase coverage of the light curve). Since the system is out of the plane, we assume that the full foreground extinction applies. The value given by Schlafly \& Finkbeiner (2011) is $E(B-V) = 0.10$. Thus we find  $B_0 = 16.76$ and $V_0 = 15.91$, so $(B-V)_0 \sim 0.85$. The day side color ($\phi = 0.5$) is $(B-V)_0 \sim 0.90$, which is slightly redder as expected due to gravity darkening. 

\begin{figure}[ht]
\includegraphics[width=3.3in]{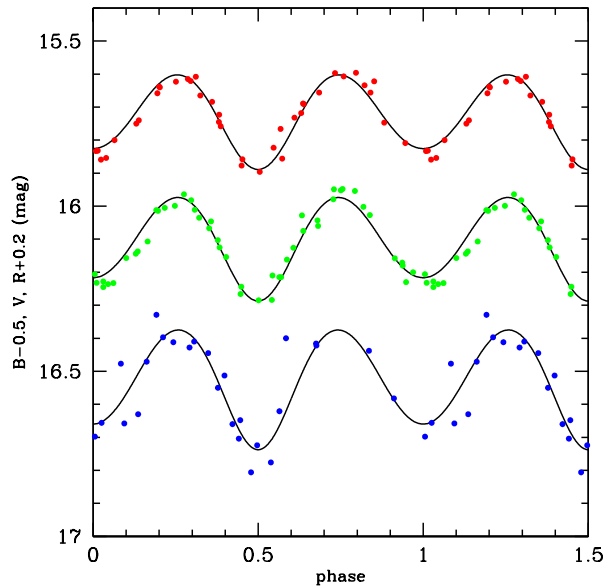}
\caption{PROMPT photometry in $BVR$ (blue, green, and red points) plotted against ellipsoidal models assuming $i = 58^{\circ}$ and $T_{\rm eff}  = 5000$ K. The photometry as plotted is not corrected for foreground reddening.}
\end{figure}

Using the color--temperature relation of Sekiguchi \& Fukugita (2000) and a metallicity of [Fe/H] = --0.65, the night side $(B-V)_0$ color suggests $T_{\rm eff} \sim 5000$ K (it would be about 200 K lower for solar metallicity). This is consistent with the $T_{\rm eff}$ inferred from a Marigo \etal~(2008) 10 Gyr isochrone of the same metallicity for this $(B-V)_0$. Given the unusual evolutionary state of the star with a significant amount of mass loss (see below), this value should only be taken as a rough estimate.

\subsection{Evidence for Accretion}

As discussed in \S 2.5.1, \aJ1417 shows persistent double-peaked H$\alpha$ emission that is evidence for an accretion disk in the system. The H$\alpha$ emission varies in its equivalent width and line profile, displaying both secular and orbital changes. The top panel of Figure 4 shows H$\alpha$ emission line profiles corrected to the rest frame of the secondary, averaged over each observing epoch, and both normalized and scaled to exhibit variations in the profile shape. There are 13 distinct epochs plotted in Figure 4 (as discussed in \S 2.5.1, a subset of the observations did not include H$\alpha$).

A minimum is consistently seen in the line profile at the systemic velocity of the secondary (adjusted here to 0 km s$^{-1}$ at all epochs; dashed line in Figure 4). In many observations, the emission line is clearly double-peaked---a classic signature of an accretion disk. The emission appears to be preferentially blueshifted in the first half of the orbital period ($\phi$ = 0.0--0.5) and redshifted in the latter half, although we see a few strong exceptions to this trend (e.g., the spectrum from 2013 Jan 15.4).

The spectrum from 2014 Jun 15.2 ($\phi = 0.12$) shows nearly symmetric double-peaked emission, which we take as a reference. At this epoch the two peaks are separated by 234 km s$^{-1}$ (marked with dotted lines in Figure 4; the blue peak is centered at $-112\pm5$ km s$^{-1}$ and the red at $122\pm5$ km s$^{-1}$). The location of these peaks is consistent at many other epochs, although additional components with more extreme velocities are sometimes superimposed or dominant. 

The observed peak separation of 234 km s$^{-1}$ is surprising if we take the simple interpretation of an accretion disk such that half the separation (117 km s$^{-1}$) is the radial velocity of the outer disk (Smak 1981). Within the uncertainties this value is identical to the orbital semi-amplitude ($K_2 = 115$ km s$^{-1}$), implying the disk entirely fills the space between the stars. This puzzle is reminiscent of that in the black hole binary V404 Cyg, which also has a red giant companion. In this system the peak separation of the H$\alpha$ emission components and $K_2$ are similar to those observed in \aJ1417. Despite intensive study, in V404 Cyg the emission profile has not been adequately explained, except to conclude that the H$\alpha$ emission is unlikely to be explained as a simple Keplerian disk (Casares et al.~1993).

We show variations in the H$\alpha$ equivalent width as a function of orbital phase in the bottom panel of Figure 4. Variability is observed on timescales of hours, especially on 2015 Feb 18, when the strength and velocity width of the emission both $\sim$ double during the night.

There is evidence of orbital variation, with lower equivalent width around $\phi \sim 0.5$ (compared to $\phi \sim 0$), but this potential signal is contaminated by the observed secular changes in the H$\alpha$ emission. Similar hints of H$\alpha$ orbital modulation are seen in V404~Cyg (Casares et al.~1993). We note that many spectra were not obtained in photometric conditions and are therefore not flux-calibrated in an absolute sense; if the optical continuum varies with the H$\alpha$ emission as might be expected in an accretion disk, then equivalent width variations place a lower limit on the fluctuation of H$\alpha$ relative to the companion star flux (e.g., Hynes et al.~2002).

\begin{figure}[ht]
\includegraphics[width=3.3in]{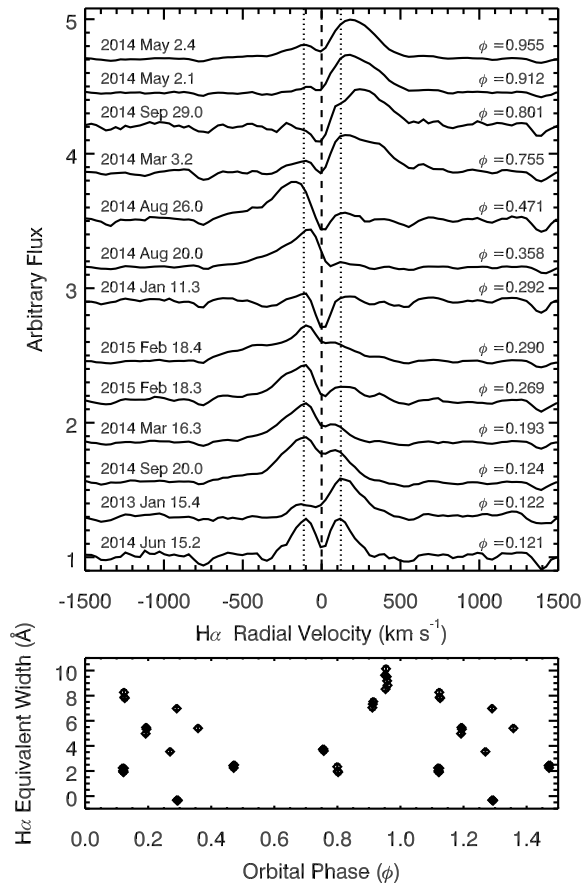}
\caption{{\bf Top:} H$\alpha$ profile as a function of orbital phase (though covering many orbits). The UVES spectrum (2014 Aug 20) has been smoothed to match the resolution of the SOAR spectra. {\bf Bottom:} Equivalent width of the H$\alpha$ emission as a function of orbital phase.}
\end{figure}

The strongest H$\alpha$ emission is observed on 2014 May 2 ($\phi = 0.91$ and 0.95), accompanied by a redshifted profile. These spectra are strongly reminiscent of the redshifted H$\alpha$ ``flares" observed in V404 Cyg, which may signify a outflow from the accretion disk or central source which absorbs the blue wing of the line. In that system these variations are hypothesized to arise from magnetic reconnection events or variable irradiation of the accretion disk (Hynes et al. 2002). 

Overall, we interpret the H$\alpha$ emission as evidence of an accretion disk, with rich phenomenology that deserves further study. In their ``high" low-mass X-ray binary states, all three transitional millisecond pulsar systems show H$\alpha$ emission in accretion disks (Wang et al.~2009; Pallanca et al.~2013; de Martino et al.~2014). In the globular cluster NGC 6397, the companion of PSR J1740-5340A shows H$\alpha$ emission associated with an outflow rather than a disk (Sabbi et al.~2003). This star is a stripped giant (Bogdanov et al.~2010; Mucciarelli et al.~2013) and so presents an interesting parallel to \aJ1417 (see \S 4) with an evolved companion but in a rotational-powered rather than accretion-powered state.




\subsection{Ellipsoidal Variations and Inclination}

The CSS photometry shows two minima and maxima per period, suggesting the presence of ellipsoidal variations due to tidal deformation of the secondary. Modeling the ellipsoidal variations can constrain the inclination of the binary, with the largest effect at edge-on inclinations ($i=90\deg$) and no ellipsoidal variations at face-on inclinations ($i=0\deg$).

However, the very broad CSS filter and moderately large uncertainties in the photometry are not ideal for modeling the light curve. Therefore, we use the less well-sampled but more precise PROMPT photometry in $BVR$ for modeling the light curve.

We make several assumptions in the light curve modeling. First, given the evidence for an accretion disk, we assume the secondary fills its Roche lobe. We also assume that the accretion disk does not contribute significantly to the optical luminosity of the system (this is justified by the results below). Due to the modest X-ray luminosity of the system ($\sim 10^{33}$ erg s$^{-1}$; see \S 2.2) and long period, irradiation of the secondary is expected to be negligible: the X-ray flux at the stellar surface is $\lesssim 10^8$ erg cm$^{-2}$ s$^{-1}$, and the ratio of optical to X-ray flux at the surface of the star is $\gtrsim 100$; indeed, we see no evidence for irradiation in the light curve. Finally, we fix the values of the period and mass ratio to those determined via spectroscopy.

 
We model the light curve using  {\tt XRBinary}\footnote{http://www.as.utexas.edu/$\sim$elr/XRbinaryV2.4/} (Bayless \etal~2010). Given the assumptions above, the only free parameters in our fits are the inclination, $T_{\rm eff}$, and overall normalization of the model.

Given the results of \S 3.2, we begin by fitting models with $T_{\rm eff}  = 5000$ K. Photometry from each band was modeled independently. The best-fit inclinations using the $B$, $V$, and $R$ photometry are $i = 59\pm3^{\circ}$, $58\pm2^{\circ}$, and $58\pm2^{\circ}$. These are clearly in excellent agreement. The listed uncertainties in the inclinations are statistical only.

The ellipsoidal models for $BVR$ are plotted against the data in Figure 3. We note that the models are not perfect representations of the data; for example, at $\phi$ between $\sim 0$ and 0.2, the data appear to slightly lag the model, though there is no corresponding feature at any other phase. The phase coverage of the data is also incomplete, especially in $B$. The light curve modeling should be revisited in the future with improved photometric data.

To show the effect of changing the temperature, fits were also performed with the $T_{\rm eff}$ increased or decreased by 250 K. These changes had a minor systematic effect on the best-fit inclinations, with $i$ increasing by 1--$1.5^{\circ}$ at $T_{\rm eff} = 5250$ K and decreasing by the same amount for $T_{\rm eff} = 4750$ K. These give a sense of the systematic uncertainties due to $T_{\rm eff}$ on modeling the ellipsoidal variations.

Another possible systematic error in the photometric modeling is the ``veiling" of the secondary if there is a significant contribution from disk light. Qualitatively, veiling leads to an underestimate of the inclination as the amplitude of the ellipsoidal modulations is suppressed (e.g., Casares et al.~1993). That the same inclination is derived for each of $BVR$ implies either that veiling is not important, or that the effective temperature of the emitting region of the disk is the same as that of the secondary. Future near-IR observations could improve constraints on a possible disk contribution to the light curve, though veiling may be present even toward redder wavelengths in low-mass X-ray binaries (e.g., Kreidberg et al.~2012).

For the remainder of the paper we adopt  $i = 58\pm2^{\circ}$. 

\subsection{Component Masses}

With the measured spectroscopic semi-amplitude and period, mass ratio from the rotational broadening, and inclination from photometry, we have enough information to determine both the primary and secondary masses in the system. Our estimate of the primary mass is $M_1 = 1.97\pm0.15 M_{\odot}$, consistent with a massive neutron star. The secondary is $0.35\pm0.04 M_{\odot}$. 

We emphasize that the uncertainties on these values are solely statistical and do not consider the systematic uncertainties described above in the derivation of the inclination. However, a substantial increase in the inclination would be necessary to reduce the inferred primary mass to a value consistent with a white dwarf rather than a neutron star: $i =72^{\circ}$ would imply a primary mass of $1.4 M_{\odot}$. Given the measured mass ratio, the minimum primary mass allowed is only $1.2 M_{\odot}$, so the permitted mass range of a putative white dwarf primary is limited. In addition, a symbiotic star (accreting white dwarf with giant secondary) with a period of 5.4 d would be extraordinarily unusual---most have periods of at least hundreds of days (Belczy{\'n}ski et al.~2000). Future X-ray observations can provide additional constraints on the inclination; depending on the geometry of the accretion disk, inclinations above $i > 70^{\circ}$ would produce X-ray dips or eclipses. 


\subsection{Distance and Kinematics}

The normalization of the best-fit light curve model gives the mean bolometric luminosity of the secondary. Using the same Marigo \etal~(2008) isochrone discussed above, the bolometric correction to $V$ is 0.23 mag. Given an observed mean $V$ mag of $\overline{V_0} = 15.80$, the implied distance is 4.4 kpc. The principal uncertainty in this estimate is systematic, and realistically the distance is likely uncertain by at least 20\%.

The Galactic coordinates of \aJ1417 are ($l \sim 318.9^{\circ}$, $b \sim 16.1^{\circ}$): the system is in the direction of the inner Galaxy, but above the Plane. The latitude corresponds to a height of $\sim 1.2$ kpc above the Plane for a distance of 4.4 kpc.

Using this distance, the proper motion of \aJ1417 (\S 2.4), and the radial velocity derived in \S 3.1, we can derive the 3-D space motion of the system, which is ($U$,$V$,$W$) = ($141\pm74$, --$137\pm82$, --$2\pm96$) km s$^{-1}$ (for this estimate, a $1\sigma$ distance uncertainty of  0.5 kpc was assumed). While the uncertainties are large due to the corresponding substantial uncertainty in the optical proper motion, the 3-D velocity of the binary ($219_{-92}^{+103}$ km s$^{-1}$) is consistent with that observed for other millisecond pulsars (e.g., Gonzalez et al.~2011). Given this high velocity, the system's current spatial location may not reflect its origin; nonetheless, the modest evidence for subsolar metallicity in \S 3.2 would also be more consistent with an origin in the thick disk than the thin disk. 

The properties of the system are summarized in Table 1.

\subsection{A Chance Alignment?}

Here we briefly consider whether it is plausible that the low-mass X-ray binary described in this paper is not associated with the \emph{Fermi}-LAT source that prompted its discovery. The total
Galactic population of low-mass X-ray binaries is estimated to be $\sim 10^4$ (Jonker et al.~2011), with most located in the disk or bulge; of the remainder, some will be more distant than the 4.4 kpc estimated for this source, or too faint to detect in X-rays. Even optimistically assuming a population of $\sim 1000$ low-mass X-ray binaries that meet these constraints, the space density would only be about 1 per 30 deg$^{2}$, giving a chance alignment probability of $\lesssim 10^{-4}$ within the 95\% \emph{Fermi}-LAT error ellipse of \aJ1417. Even this estimate is sanguine, as the X-ray binary is located $< 25\arcsec$ from the 3FGL \emph{Fermi}-LAT centroid. We conclude that a chance alignment between the low-mass X-ray binary and  \aJ1417 is very unlikely. 

\section{Discussion}

Using spectroscopy and photometry we have shown that the \emph{Fermi} $\gamma$-ray source \aJ1417 is likely associated with a low-mass X-ray binary, whose primary is likely a neutron star of nearly $2 M_{\odot}$. Given the low mass of the hydrogen-rich giant secondary ($\lesssim 0.4 M_{\odot}$), it likely followed a standard evolutionary path for a low-mass X-ray binary with a neutron star primary, initiating Case B mass transfer after leaving the main sequence. This indicates the system is likely to evolve into a standard millisecond pulsar with a He white dwarf companion at the end of the accretion phase. 

Tauris \& Savonije (1999) argue that much of the mass in the accretion flow in neutron star binaries with giant secondaries is lost rather than accreted onto the neutron star. In this case, the inferred high mass of the neutron star implies that it was born with a mass above the standard $1.4 M_{\odot}$. A similar conclusion has been reached for PSR J1614$-$2230, which has an accurate pulsar mass of $1.97\pm0.04 M_{\odot}$ measured via the Shapiro delay (Demorest \etal~2010), though in this case the secondary is a CO white dwarf that likely evolved from an intermediate-mass X-ray binary (Tauris et al.~2011). Heavy neutron star masses have been inferred for some millisecond pulsars discovered via \emph{Fermi} $\gamma$-ray emission (Romani et al.~2012; Schroeder \& Halpern 2014).

Very few low-mass X-ray binaries that are actively accreting have been detected as \emph{Fermi}-LAT sources (Acero et al.~2015), with the exception of the two of the three transitional millisecond pulsars (de Martino et al.~2010; Stappers et al.~2014). In addition, Bogdanov \& Halpern (2015) have shown that the \emph{Fermi} source 3FGL J1544.6--1125 is associated with an X-ray binary that exhibits optical and X-ray properties similar to the other transitional millisecond pulsars. The consistent presence of $\gamma$-ray emission among accreting transitional millisecond pulsars motivates us to suggest that  \aJ1417 may also be a transitional millisecond pulsar. There is no evidence for a state change in \aJ1417 over the last $\sim 10$ years: the CSS photometry shows a constant mean magnitude since 2005 (\S 2.4.1), while PSR J1023+0038 became brighter by $\sim 1$ mag during its recent change to the disk state (Halpern et al.~2013).

The origin of the $\gamma$-ray emission in \aJ1417 is uncertain. For most millisecond pulsars the $\gamma$-ray emission is thought to come from the magnetosphere of the pulsar and be entirely unrelated to the secondary. The lack of radio emission and evidence for an accretion disk in this system suggests that this is unlikely to be the correct explanation in \aJ1417. Indeed, in the transitional system PSR J1023+0038 the $\gamma$-ray emission increased by a factor of at least 5 during the recent state change when the pulsar became undetectable (Stappers et al.~2014). They hypothesized that the pulsar was enshrouded but still active, with the increase in $\gamma$-ray flux coming from a shock between the pulsar wind and surrounding material, related to the re-appearance of an accretion disk. However, Archibald et al.~(2015) show that J1023+0038 exhibits coherent X-ray pulsations in the high state that indicate accreted material is reaching the neutron star. This material ought to quench the pulsar, so it is not trivial to associate the $\gamma$-ray emission with a pulsar wind. An alternative model is that the $\gamma$-rays are due to the interaction between a relativistic jet and the surrounding material, or to synchrotron self-Compton radiation produced at the boundary of the inner accretion disk (Papitto et al.~2014; Deller et al.~2015).

The X-ray luminosity of \aJ1417 ($\sim 1.4 \times 10^{33}$ erg s$^{-1}$) is consistent with the average value observed in the disk state of the known transitional millisecond pulsars (Linares 2014), and could well suggest that mode switching is also occurring in \aJ1417, but is not resolved in our short X-ray observations due to the greater distance and hence lower flux of this source. The X-ray photon index ($\Gamma \sim 1.3$) is also consistent with the transitional objects in this state. The ratio of $\gamma$-ray to X-ray luminosity is $\sim 20$, which is substantially larger than that observed in the disk state for J1023+0038, perhaps related to the giant companion.


The transitional millisecond pulsars, including J1023+0038 (Deller et al.~2015), have been detected as flat-spectrum cm radio continuum sources in the low-mass X-ray binary state. The observed 10 GHz radio continuum flux density of J1023+0038 would correspond to $\sim 5$--10 $\mu$Jy at the inferred distance of \aJ1417, which is marginally detectable with southern radio interferometers.

Clearly, additional work is needed to characterize \aJ1417. Continuing observations with \emph{Fermi}-LAT could allow a measurement of $\gamma$-ray variability of the source, which would help determine the origin of the $\gamma$-ray emission. Radio observations at superior conjunction should be used to search for radio pulsations or provide strong constraints on their presence, although radio pulsar emission is not expected while the source has an accretion disk. Improved photometric data and modeling could aid in constraining the systematic uncertainties on the mass estimate of the neutron star. 
Observations of X-ray variability on short timescales, an improved X-ray spectrum, and the detection of radio continuum emission offer the best current hope for determining whether  \aJ1417 is a transitional millisecond pulsar.




\begin{deluxetable}{lr}
\tablecaption{Summary of Properties  \label{tab:dat2}}
\startdata
Opt. R.A. (J2000 h:m:s)  &  14:17:30.60 \\ 
Opt. Dec. (J2000 $^\circ$ : \arcmin : \arcsec) &  --44:02:57.4 \\
Period (d)      & $5.37385\pm0.00035$ \\
$K_2$ (km s$^{-1}$) &  $115.7\pm1.1$ \\ 
$T_{0.5}$ (d) &  $2457067.605\pm0.024$ \\
$V_{rot}$ sin $i$ (km s$^{-1}$)  & $33.6\pm0.7$ \\
$M_2/M_1$ & $0.179\pm0.010$ \\
$i$ ($^{\circ}$) & $58\pm2$ \\
$a$ ($R_{\odot}$) & $17.1\pm0.4$ \\
$V_0$, $\phi = 0$ (mag) & 15.91\\
$(B-V)_0$, $\phi = 0$ (mag) & 0.85\\
$M_1$ ($M_{\odot}$) & $1.97\pm0.15$\\
$M_2$ ($M_{\odot}$) & $0.35\pm0.04$\\
$v_{sys}$  (km s$^{-1}$)  &  --$15.3\pm0.9$ \\
$\mu_{\alpha} \textrm{cos} \, \delta$ (mas yr$^{-1}$ ) & --$8.8\pm5.0$ \\
$\mu_{\delta}$ (mas yr$^{-1}$ ) &  --$3.1\pm4.7$ \\
$(U,V,W)$ (km s$^{-1}$)  & ($141\pm74$, --$137\pm82$, --$2\pm96$)\\
Distance (kpc) & 4.4 \\
\enddata
\end{deluxetable}

\acknowledgments

We thank an anonymous referee for comments that significantly improved the paper. We thank the Aspen Center for Physics and the NSF Grant \#1066293 for hospitality during the writing of this paper. We thank P.~Groot, T.~Tauris, T.~Maccarone, S.~Bogdanov, and C.~Heinke for useful conversations, and M.~Coe and G.~McSwain for early discussions on possible follow-up. Work by C.C.C. at NRL is supported in part by NASA DPR S-15633-Y.

Based on observations obtained at the Southern Astrophysical Research (SOAR) telescope, which is a joint project of the Minist\'{e}rio da Ci\^{e}ncia, Tecnologia, e Inova\c{c}\~{a}o (MCTI) da Rep\'{u}blica Federativa do Brasil, the U.S. National Optical Astronomy Observatory (NOAO), the University of North Carolina at Chapel Hill (UNC), and Michigan State University (MSU). Also based on observations made with ESO telescopes at the La Silla Paranal Observatory under programme ID 293.D-5029. The scientific results reported in this article are based in part on data obtained from the Chandra Data Archive. This paper includes archived data obtained through the Australia Telescope Online Archive. The CSS survey is funded by the National Aeronautics and Space Administration under Grant No.~NNG05GF22G issued through the Science Mission Directorate Near-Earth Objects Observations Program. The CRTS survey is supported by the U.S.~National Science Foundation under grants AST-0909182 and AST-1313422.

{}


\begin{thebibliography}{}

\bibitem[Abdo et al.(2010)]{1fgl} Abdo, A.~A., Ackermann, M., Ajello, M., et al.\ (\Fermi-LAT collaboration) 2010, \apjs, 188, 405 (1FGL catalog)
\bibitem[Acero et al.(2015)]{3fgl} Acero, F., et al.\ (\Fermi-LAT collaboration) 2015, ApJS, submitted (arXiv: 1501.02003; 3FGL catalog)
\bibitem[Ackermann et al.(2015)]{2015arXiv150106054A} Ackermann, M., Ajello, M., Atwood, W., et al.\ 2015, ApJ, submitted (arXiv:1501.06054)
\bibitem[Archibald et al.(2009)]{2009Sci...324.1411A} Archibald, A.~M., Stairs, I.~H., Ransom, S.~M., et al.\ 2009, Science, 324, 1411 
\bibitem[Archibald et al.(2013)]{2013arXiv1311.5161A} Archibald, A.~M., Kaspi, V.~M., Hessels, J.~W.~T., et al.\ 2013, ApJ, submitted (arXiv:1311.5161)
\bibitem[Archibald et al.(2015)]{2014arXiv1412.1306A} Archibald, A.~M., Bogdanov, S., Patruno, A., et al.\ 2015, ApJ, submitted (arXiv:1412.1306)
\bibitem[Bassa et al.(2014)]{2014MNRAS.441.1825B} Bassa, C.~G., Patruno, A., Hessels, J.~W.~T., et al.\ 2014, \mnras, 441, 1825 
\bibitem[Baumgartner et al.(2013)]{2013ApJS..207...19B} Baumgartner, W.~H., Tueller, J., Markwardt, C.~B., et al.\ 2013, \apjs, 207, 19
\bibitem[Bayless et al.(2010)]{2010ApJ...709..251B} Bayless, A.~J., Robinson, E.~L., Hynes, R.~I., Ashcraft, T.~A., \& Cornell, M.~E.\ 2010, \apj, 709, 251 
\bibitem[Belczy{\'n}ski et al.(2000)]{2000A&AS..146..407B} Belczy{\'n}ski, K., Miko{\l}ajewska, J., Munari, U., Ivison, R.~J., \& Friedjung, M.\ 2000, \aaps, 146, 407 
\bibitem[Bogdanov et al.(2010)]{2010ApJ...709..241B} Bogdanov, S., van den Berg, M., Heinke, C.~O., et al.\ 2010, \apj, 709, 241 
\bibitem[Bogdanov \& Halpern(2015)]{2015arXiv150301698B} Bogdanov, S., \& Halpern, J.~P.\ 2015, arXiv:1503.01698 
\bibitem[Bogdanov et al.(2015)]{2014arXiv1412.5145B} Bogdanov, S., Archibald, A.~M., Bassa, C., et al.\ 2015, \apj, submitted (arXiv:1412.5145)
\bibitem[Casares et al.(1993)]{1993MNRAS.265..834C} Casares, J., Charles, P.~A., Naylor, T., \& Pavlenko, E.~P.\ 1993, \mnras, 265, 834 
\bibitem[Casares (2001)]{cas01} Casares, J. 2001 {\it Binary Stars: Selected Topics on Observations and Physical Processes}, ed. F. C. Lazaro and M. J. Arevalo (Berlin: Springer) p. 277
\bibitem[Clemens et al.(2004)]{2004SPIE.5492..331C} Clemens, J.~C., Crain, J.~A., \& Anderson, R.\ 2004, \procspie, 5492, 331 
\bibitem[Crawford et al.(2013)]{2013ApJ...776...20C} Crawford, F., Lyne, A.~G., Stairs, I.~H., et al.\ 2013, \apj, 776, 20 
\bibitem[Cutri et al.(2003)]{2003tmc..book.....C} Cutri, R.~M., Skrutskie, M.~F., van Dyk, S., et al.\ 2003, The IRSA 2MASS All-Sky Point Source Catalog, NASA/IPAC Infrared Science Archive
\bibitem[de Martino et al.(2010)]{2010A&A...515A..25D} de Martino, D., Falanga, M., Bonnet-Bidaud, J.-M., et al.\ 2010, \aap, 515, AA25 
\bibitem[de Martino et al.(2014)]{2014MNRAS.444.3004D} de Martino, D., Casares, J., Mason, E., et al.\ 2014, \mnras, 444, 3004 
\bibitem[Deller et al.(2015)]{2014arXiv1412.5155D} Deller, A.~T., Mold{\'o}n, J., Miller-Jones, J.~C.~A., et al.\ 2015, ApJ, submitted (arXiv:1412.5155)
\bibitem[Demorest et al.(2010)]{2010Natur.467.1081D} Demorest, P.~B., Pennucci, T., Ransom, S.~M., Roberts, M.~S.~E., \& Hessels, J.~W.~T.\ 2010, \nat, 467, 1081 
\bibitem[Drake et al.(2009)]{dra09} Drake, A.~J., Djorgovski, S.~G., Mahabal, A., et al.\ 2009, \apj, 696, 870
\bibitem[Eastman et al.(2010)]{2010PASP..122..935E} Eastman, J., Siverd, R., \& Gaudi, B.~S.\ 2010, \pasp, 122, 935 
\bibitem[Eggleton(1983)]{1983ApJ...268..368E} Eggleton, P.~P.\ 1983, \apj, 268, 368 
\bibitem[Fruscione et al.(2006)]{fru06} Fruscione, A., McDowell, J.~C., Allen, G.~E., et al.\ 2006, \procspie, 6270, 60
\bibitem[Gonzalez et al.(2011)]{2011ApJ...743..102G} Gonzalez, M.~E., Stairs, I.~H., Ferdman, R.~D., et al.\ 2011, \apj, 743, 102
\bibitem[Gray \& Corbally(2014)]{2014AJ....147...80G} Gray, R.~O., \& Corbally, C.~J.\ 2014, \aj, 147, 80 
\bibitem[Halpern et al.(2013)]{2013ATel.5514....1H} Halpern, J.~P., Gaidos, E., Sheffield, A., Price-Whelan, A.~M., \& Bogdanov, S.\ 2013, The Astronomer's Telegram, 5514, 1 
\bibitem[Hiroi et al.(2013)]{2013ApJS..207...36H} Hiroi, K., Ueda, Y., Hayashida, M., et al.\ 2013, \apjs, 207, 36 
\bibitem[Hynes et al.(2002)]{2002MNRAS.330.1009H} Hynes, R.~I., Zurita, C., Haswell, C.~A., et al.\ 2002, \mnras, 330, 1009 
\bibitem[Jonker et al.(2011)]{2011ApJS..194...18J} Jonker, P.~G., Bassa, C.~G., Nelemans, G., et al.\ 2011, \apjs, 194, 18 
\bibitem[Kalberla et al.(2005)]{kal05} Kalberla, P.~M.~W., Burton, W.~B., Hartmann, D., et al.\ 2005, \aap, 440, 775
\bibitem[Kreidberg et al.(2012)]{2012ApJ...757...36K} Kreidberg, L., Bailyn, C.~D., Farr, W.~M., \& Kalogera, V.\ 2012, \apj, 757, 36 
\bibitem[Landolt(1992)]{1992AJ....104..340L} Landolt, A.~U.\ 1992, \aj, 104, 340
\bibitem[Linares(2014)]{2014ApJ...795...72L} Linares, M.\ 2014, \apj, 795, 72 
\bibitem[Marigo et al.(2008)]{2008A&A...482..883M} Marigo, P., Girardi, L., Bressan, A., et al.\ 2008, \aap, 482, 883 
\bibitem[Mucciarelli et al.(2013)]{2013ApJ...772L..27M} Mucciarelli, A., Salaris, M., Lanzoni, B., et al.\ 2013, \apjl, 772, LL27
\bibitem[Monet et al.(2003)]{mon03} Monet, D.~G., Levine, S.~E., Canzian, B., et al.\ 2003, \aj, 125, 984 
\bibitem[Nolan et al.(2012)]{2012ApJS..199...31N} Nolan, P.~L., Abdo, A.~A., Ackermann, M., et al.\ 2012, \apjs, 199, 31 
\bibitem[Pallanca et al.(2013)]{2013ApJ...773..122P} Pallanca, C., Dalessandro, E., Ferraro, F.~R., Lanzoni, B., \& Beccari, G.\ 2013, \apj, 773, 122 
\bibitem[Papitto et al.(2014)]{2014MNRAS.438.2105P} Papitto, A., Torres, D.~F., \& Li, J.\ 2014, \mnras, 438, 2105 
\bibitem[Papitto et al.(2013)]{2013Natur.501..517P} Papitto, A., Ferrigno, C., Bozzo, E., et al.\ 2013, \nat, 501, 517 
\bibitem[Petrov et al.(2013)]{pet13} Petrov, L., Mahony, E.~K., Edwards, P.~G., Sadler, E.~M., \& Schinzel, F.~K.\ 2013, \mnras, 432, 1294
\bibitem[Ptak \& Griffiths(2003)]{pta03} Ptak, A., \& Griffiths, R.\ 2003, Astronomical Data Analysis Software and Systems XII, 295, 465
\bibitem[Reichart et al.(2005)]{2005NCimC..28..767R} Reichart, D., Nysewander, M., Moran, J., et al.\ 2005, Nuovo Cimento C Geophysics Space Physics C, 28, 767 
\bibitem[Roberts(2013)]{2013IAUS..291..127R} Roberts, M.~S.~E.\ 2013, IAU Symposium, 291, 127 
\bibitem[Romani et al.(2012)]{2012ApJ...760L..36R} Romani, R.~W., Filippenko, A.~V., Silverman, J.~M., et al.\ 2012, \apjl, 760, L36 
\bibitem[ROSAT Consortium(2000)]{ros00} ROSAT Consortium, 2000, The Second ROSAT Source Catalog of Pointed Observations, ROSAT News 72, 25-May-2000
\bibitem[Roy et al.(2015)]{2014arXiv1412.4735R} Roy, J., Ray, P.~S., Bhattacharyya, B., et al.\ 2015, \apjl, 800, L12
\bibitem[Ruf(1999)]{Ruf} Ruf, T.\ 2009, Biol. Rhythm Res., 30, 178
\bibitem[Sabbi et al.(2003)]{2003ApJ...589L..41S} Sabbi, E., Gratton, R., Ferraro, F.~R., et al.\ 2003, \apjl, 589, L41 
\bibitem[Schlafly \& Finkbeiner(2011)]{2011ApJ...737..103S} Schlafly, E.~F., \& Finkbeiner, D.~P.\ 2011, \apj, 737, 103 
\bibitem[Schinzel et al.(2015)]{2015ApJS..217....4S} Schinzel, F.~K., Petrov, L., Taylor, G.~B., et al.\ 2015, \apjs, 217, 4 
\bibitem[Schroeder \& Halpern(2014)]{2014ApJ...793...78S} Schroeder, J., \& Halpern, J.\ 2014, \apj, 793, 78 
\bibitem[Sekiguchi \& Fukugita(2000)]{2000AJ....120.1072S} Sekiguchi, M., \& Fukugita, M.\ 2000, \aj, 120, 1072 
\bibitem[Stappers et al.(2014)]{2014ApJ...790...39S} Stappers, B.~W., Archibald, A.~M., Hessels, J.~W.~T., et al.\ 2014, \apj, 790, 39 
\bibitem[Smak(1981)]{1981AcA....31..395S} Smak, J.\ 1981, AcA, 31, 395 
\bibitem[Strader et al.(2014)]{2014ApJ...788L..27S} Strader, J., Chomiuk, L., Sonbas, E., et al.\ 2014, \apjl, 788, L27 
\bibitem[Tauris \& Savonije(1999)]{1999A&A...350..928T} Tauris, T.~M., \& Savonije, G.~J.\ 1999, \aap, 350, 928 
\bibitem[Tauris \& van den Heuvel(2006)]{2006csxs.book..623T} Tauris, T.~M., \& van den Heuvel, E.~P.~J.\ 2006, Compact stellar X-ray sources, 623 
\bibitem[Tauris et al.(2011)]{2011MNRAS.416.2130T} Tauris, T.~M., Langer, N., \& Kramer, M.\ 2011, \mnras, 416, 2130 
\bibitem[Voges et al.(1999)]{vog99} Voges, W., Aschenbach, B., Boller, T., et al.\ 1999, \aap, 349, 389
\bibitem[Wang et al.(2009)]{2009ApJ...703.2017W} Wang, Z., Archibald, A.~M., Thorstensen, J.~R., et al.\ 2009, \apj, 703, 2017 
\bibitem[Wang et al.(2012)]{2012ApJ...761...46W} Wang, S.~X., Wright, J.~T., Cochran, W., et al.\ 2012, \apj, 761, 46 
\bibitem[Zacharias et al.(2013)]{2013AJ....145...44Z} Zacharias, N., Finch, C.~T., Girard, T.~M., et al.\ 2013, \aj, 145, 44 



\end{thebibliography}
\end{document}